# Performance Evaluation of Orchestra Scheduling in Time-Slotted Channel Hopping Networks

Reza Amini
Faculty of Engineering, Vrije Universiteit Brussel
Brussel, Belgium
reza.amini@vub.be

Mehdi Imani
IEEE Member
Stockholm, Sweden
m.imani@gmail.com

Petio Ivanov Todorov
Faculty of Engineering, Vrije Universiteit Brussel
Brussel, Belgium
petio.ivanov.todorov@vub.be

Maaruf Ali
Department of Computer Engineering, Epoka University
Vorë, Tirana, Albania
maaruf@ieee.org

*Abstract*— In this paper, we evaluate the performance of networks that use RPL (Routing Protocols for Low Power and Lossy Networks) with TSCH (Time Slotted Channel Hopping) and Orchestra (an autonomous method for building the TSCH schedule). We measure the performance in the transient state when a node dies (i.e., removed from the network) and determine how long it takes for the network to come back to a stable RPL tree and also what the impact is with respect to energy consumption. Our analysis shows that the Orchestra reduces the energy consumption when the RPL is in a transient state, like in the case of when one of the nodes die. Furthermore, we calculate the energy consumption in the transient state without using Orchestra, and then we make a comparison between both outcomes. We show that Orchestra reduces energy consumption by up to one-third compared to not using Orchestra.

## I. INTRODUCTION

Low Power and Lossy Wireless Sensor Networks (WSN) are part of the IoT as they can be used to collect data and to actuate events in networks remotely over the internet. Generally, the devices connected to these networks are constrained in terms of memory and power. Since these networks can be quite large, requiring many devices, the devices must be low cost and hence have small memory. Moreover, these devices should consume very little energy so that they can be used for a long time without any manual maintenance. These constraints have created a new set of challenges that researchers are trying to meet. Much research has been conducted in this area to minimize the power consumption and maximize the network lifetime both in synchronous and asynchronous mode [1-8]. The IEEE standard 802.15.4e [9] defines several MAC layer amendments to support industrial application domains. Currently, these amendments have been incorporated in the IEEE standard 802.15.4-2015 [10]. Time Slotted Channel Hopping (TSCH) is one of them that combines several features like time-slot, channel hopping, and multiple channels. The rest of this paper is organized as follows: in Section Two, the related works are reviewed. In Sections Three and Four, an overview of TSCH and Orchestra are presented. In Section Five, different scenarios are evaluated, and the results are compared. Finally, Section Six presents the conclusions of the paper.

## II. RELATED WORK

There are several studies that have already evaluated the performance of RPL topology under steady-state. Recently, some studies have evaluated the performance of RPL under a transient state like when one of the nodes die, and the tree has to be rebuilt in order to come back to a steady-state again. In such a case, TSCH, as well as Orchestra, play an important role in reducing the energy consumption as much as possible while the tree is rebuilding.

Authors in [11] evaluated the performance of RPL in two states without using TSCH and Orchestra. It showed that the transient state leads to a significant decrease in performance compared to the steady-state case, particularly when changes frequently occurred in the RPL topology. Authors in [12] presented the impacts of using time-slotted channel hopping on the performance of RPL. They explained that there are several time schedules, including centralized and distributed. Their work proved that TSCH leads to more reliability due to channel hopping with different frequencies and a guaranteed bandwidth due to time-slotting. Authors in [13] investigated the impacts of using Orchestra scheduling through TSCH on the performance of RPL. They demonstrated that Orchestra made the performance in the transient state more robust. Orchestra maintains three different schedules each being allocated to a particular traffic plan including the application traffic, RPL traffic and for the TSCH beacon.

## III. OVERVIEW OF TSCH

TSCH is a MAC-level protocol that is commonly used in WSN. TSCH creates a time-based scheme through which the nodes in the network know whether to transmit, receive, or sleep at a specific point in time [13]. As Figure 1 shows, time is divided into intervals called time-slots, and these time-slots are grouped together to form slot-frames [13]. In Figure 1, the slot-frame has a length of four which means it has four 4 equal length time-slots. Typically, each time-slot has a duration of ~10 ms, which is enough for one node to transmit a message and for the receiving node to send back an acknowledgment [13]. Besides the Time Slotted aspect of TSCH, there is a supplementary part that is called channel hopping. The operating bandwidth of the network is divided into physical channels; each has a different frequency. Therefore, each time-slot is divided into 16 channel offsets (i.e., logical channel numbers). The frequency that is used on a particular channel offset at a particular time-slot change with every single slot-frame. A channel offset loops through every possible frequency before it returns to the current frequency. Equation 1 [14] shows this relationship; note that V is the vector containing some channels that is available to beused by the network, ASN stands for Absolute Slot Number, and it is a basic counter for the number of time-slots that have transpired since the network was initiated, $N_{ch}$ is the length of V, *chOf* stands for channel offset, and *mod* stands for modulus operation.

$$f = V\{(ASN + chOf) \bmod N_{ch}\} \tag{1}$$

As an example, if *ASN ={4, 11, 18, 24}, chOf={1}*, and *V={16, 17, 23, 18, 26, 15, 25, …, 22, 19, 11, 12, 13, 24, 14, 20, 21} then: Nch=16, $f = V\{(4 + 1) \bmod 16\} = V[5] = 15$*, So the next channel number must be 15, shown in Figure 1.

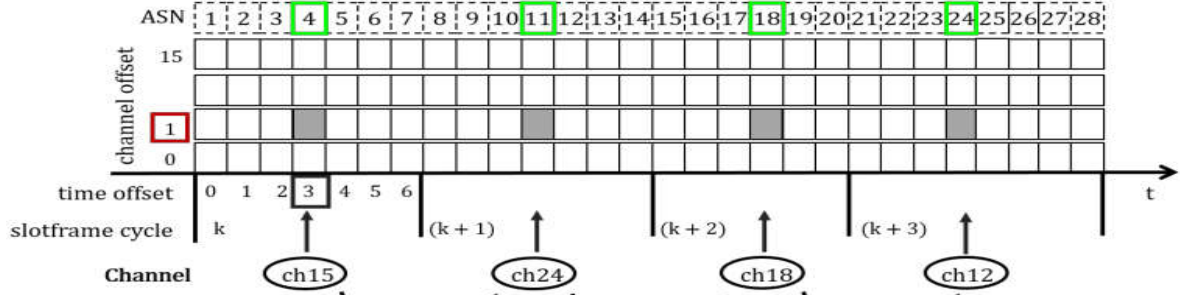

Fig. 1. An example of channel hopping in a TSCH schedule [16]

The Time Slotted aspect of TSCH guarantees bandwidth and results in a more predictable latency. The main advantage of the channel hopping aspect of TSCH is that it reduces interference. This is a result of the fact that not all nodes are active during each time-slot. Even if multiple nodes are active during the same time-slot, they can communicate on a different channel frequency, thereby avoiding interference. Furthermore, the change in frequency in the same channel/time-slot cell of each slot-frame reduces the number of potential retransmissions. Even if there is a particular constant external interference, during the next slot-frame, a node on a particular channel offset will transmit/receive on a different channel frequency, thereby avoiding the original external interference. As a result, the reliability of the network is higher with channel hopping.

The time-slot/channel offset matrix is a set of links, that is, pairs of nodes that communicate with each other. This schedule dictates which particular pairs of nodes communicate on which particular channel offset and time-slot. Dedicated links have only one pair of nodes communicating in a particular time-slot/channel offset cell, while a shared link has multiple pairs of nodes in the same cell [15]. Note that in the shared links, the pairs of nodes are communicating on the same frequency, which allows for collisions.

The method for creating this time-slot/channel offset matrix schedule is also a challenge. Generally, there are two types of schedules: centralized and distributed. In centralized scheduling, the schedule is created by a master node (usually root) after it has received topology/traffic information from all other nodes in the network [15]. When there is a change in the network, the master node recomputes the schedule and retransmits it to each of the nodes. This approach is not an ideal scheduling method for networks that are dynamic and large scale because there may be frequent changes that require the master node to constantly recompute and retransmit the schedule, thereby increasing downtime and power consumption [15]. The other type of scheduling method is the distributed method, where each individual node computes its own local schedule based on interactions with neighboring nodes. In this case there is no master node, so the energy consumption in creating the schedule is lower because it does not have to be retransmitted to all the other nodes. In both centralized and distributed scheduling, there is additional overhead because the nodes need to exchange scheduling information, in addition to network and traffic information [17].

The third alternative scheduling method is Orchestra. This scheduling method is neither centralized nor distributed, but is instead autonomous, meaning that each node "builds its own schedule without any negotiation with its neighbors" [17]. The schedule for each node is based on the RPL (IPv6 Routing Protocol for Low Power and Lossy Networks [18]) messages, which are transmitted independently of the messages for determining the TSCH schedule. The RPL messages are used for the formation and upkeep of the RPL network. As such, because the TSCH schedule is built from the RPL network/traffic information, there is no additional overhead for its creation, which should result in reduced power consumption.

IV. OVERVIEW OF ORCHESTRA

The Orchestra scheduling method [13] is used to achieve a highly-reliable and low-power TSCH system. In Orchestra, each node keeps multiple schedules and computes its own local schedules automatically based on its RPL neighbors (parents and children) in the RPL topology. The nodes allocate each of their schedules to a particular traffic plan, including application, routing, and MAC. An Orchestra schedule contains different slot-frames with different lengths [13], as shown in Figure 2. Slot-frames consists of a set of slots, with properties defined by simple scheduling rules. The slot frames repeat at specific periods, ensuring they cycle independently. If slots from different slot frames overlap, the slot in the highest priority slot frame takes precedence. The length of a slot-frame can determine traffic capacity, network latency, and energy consumption. Shorter slot-frames have their slots repeat more often, resulting in higher traffic capacity. Also, nodes have to wake up more to listen or transmit, resulting in higher energy consumption. Each slot frame is dedicated to a particular type of traffic [13]: TSCH beacons, RPL signaling traffic, and Application data

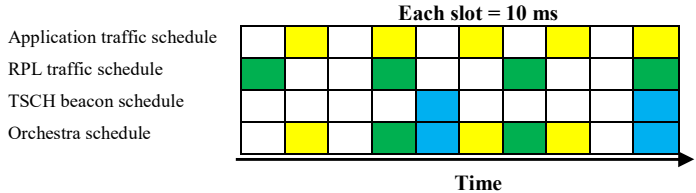

Fig. 2. An Orchestra schedule

As Figure 3 shows, there are four main types of slots in Orchestra [13]:

- Common Shared slots (CS)

These slots are used by all nodes in the network for both Tx (Transmit) and Rx (Receive) purposes.

- Receiver-Based Shared slots (RBS)

These slots are used for communication between two neighbors. At every node, an RBS slot results in one Rx slot (based on the node), and one Tx slot per neighbor (based on the neighbor).

- Sender-Based Shared slots (SBS)

These slots are similar to RBS, except that the slot coordinates are obtained from the properties of the sender node rather than the receiver. At every node, an SBS slot results in one Rx slot per neighbor (based on the neighbor) and a single Tx slot (based on the sender node).

- Sender-Based Dedicated slots (SBD)

This is a slot frame long enough to accommodate unique transmit slots to every node. These slots are similar to SBS, except they use dedicated TSCH slots instead of shared slots. Note that with TSCH dedicated slots, lost packets can be resent without using the next slot towards the same neighbor.

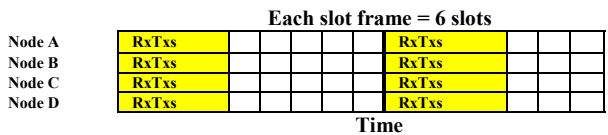

(a) Common Shared Slot

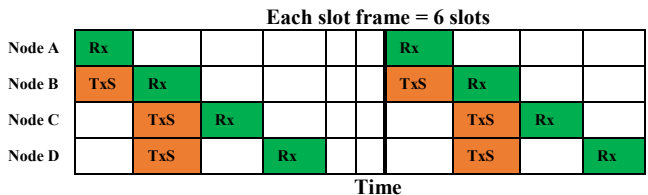

(b) Receiver-based Shared Slot

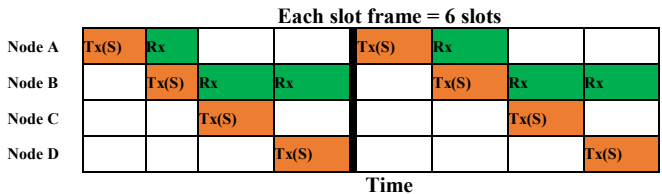

(c) Sender-based Shared Slot

Fig. 3. Different main types of slots in Orchestra

## V. PERFORMANCE ANALYSIS

We are simulating the network in the Contiki OS Java Simulator (COOJA) in the Contiki-NG distribution [19]. We have started with a tree topology. The topology uses the root, sender, and receiver nodes. First, we enabled TSCH as well as Orchestra in the make file of our simulation program. We also enabled log messages so that they are visible in the Mote Output Window. The transmission and interference ranges of each node are 50 m. The standard slot-frame length is 7 time-slots. There are 26 channels. Our current simulation network consists of 6 nodes- node #3 as the root, nodes #1, 4, 9, and 10 as receivers, and node #2 as a sender node. Figure 4 shows the layout of the network and the transmission ranges of nodes #3 and #2. Two redundant transmission paths are established between the sender, node #2, and the root, node #3. We considered this to be an interesting topology because it may be used in practice to make sure that even if one of the receiver nodes fails, there may still be an alternate route for communication to the root.

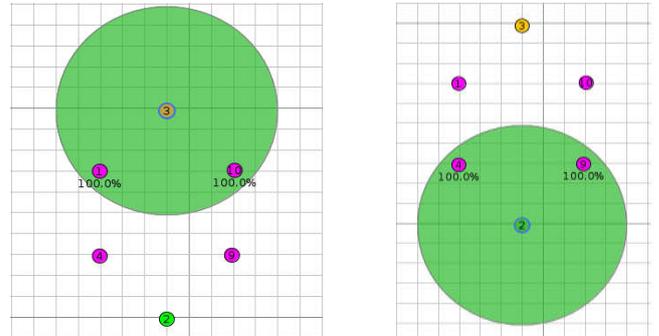

Fig. 4. The layout of the network and the transmission ranges of nodes #2 and #3.

The traffic level is dictated by sender node #2, which sends messages at a frequency of one message per second. (The nominal frequency is one message per minute, but we increased it to generate more traffic). We ran the following general experiment. on the previously mentioned topology. We paused the simulation when the network has reached a steady-state, then we removed a node and measured the power consumption until the network returned to a steady-state. In order to have a measure for comparison, we ran this experiment with Orchestra enabled on TSCH and separately without Orchestra enabled (this means we used the default scheduling method for TSCH in Cooja). Our results show that TSCH with Orchestra consume less energy in the transient state than TSCH without Orchestra because it does not require extra communication in to update the RPL network.

### A. Simulation with Orchestra enabled on TSCH

We first ran the simulation for a total of eight minutes. No motes were removed in this initial simulation. In the Mote Output window, we set the Filter to "Joined" so that we could ascertain when all nodes had joined the network. In our simulation, the last node, node #2, joined the RPL network at 41.17 seconds. From here on, we consider the amount of time the radio is ON as a representative of the power consumption.

Of course, to calculate the actual power, it is necessary to do some basic calculations involving the amount of energy that the radio uses when it is ON, the amount of time you have been running the simulation, and the Cooja PowerTracker output.

*1) Determining Steady State*

In order to determine the state of the network (steady-state or transient), we analyzed the trickle timers of each node. During the RPL network formation, the root node sends DIO (DODAG Information Objects where DODAG stands for Destination Oriented Directed Acyclic Graph) in order to inform the neighboring nodes within the transmission range. After connecting to the root node, these neighboring nodes send their own DIO messages to inform their neighboring nodes about the existing network and possibly to allow them to join. Once joined, these nodes inform their neighbors in the same manner.

DIO messages can also contain information about any changes in the network. These are useful for repairs once the network has been built. If a node dies and is removed from the network, the neighboring nodes must pass this information along. In order to make sure these messages are transmitted efficiently, RPL includes trickle timers which regulate the time interval between successive DIO messages sent by each node. If there have been no changes in the RPL network, the trickle timer interval increases so that DIO messages are sent less frequently, thus conserving energy. The interval increases from a minimum value indefinitely until it reaches some maximum value. As soon as a change is detected, a node will reset its trickle timer to the minimum value and thereby increase the frequency of messages it sends. We wrote two Python programs to parse the messages in the Cooja mote output window after saving them to a text file. We first filtered out the messages, which included the strings "trickle timer (Interval)". This way, we could monitor how the values of the trickle timer changed with time for each node. Figure 5 shows the changes in the trickle timers vs. time after running the simulation for eight minutes without removing any nodes.

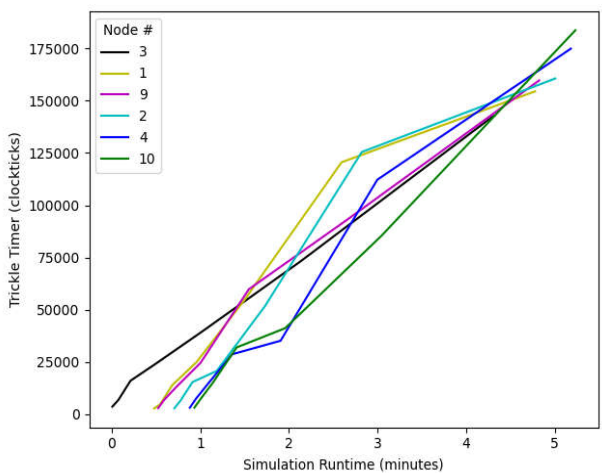

Fig. 5. The changes of the trickle timers vs. time after running the simulation for 8 minutes, without removing any nodes.

The y-axis represents the number of clock ticks of the trickle timer, where 1000 clock ticks are equal to 1 second. The x-axis is the time measured in minutes since the simulation was started. The trickle timer continues to increase steadily with time, indicating that there are no significant changes in the network.

We also created another correlated graph, Figure 6, that shows the time a particular DIO message was triggered for a particular node. We did this by filtering out the messages which ended with the string "triggered". The graph cab be read/interpret in the following way: the fourth DIO message from node #2 was triggered at approximately the 1-minute mark.

As Figure 6 shows, we can determine that the frequency of messages is decreasing with time and correspondingly, that the interval between messages is increasing. For node #3, it took approximately two minutes to send nine DIO messages.

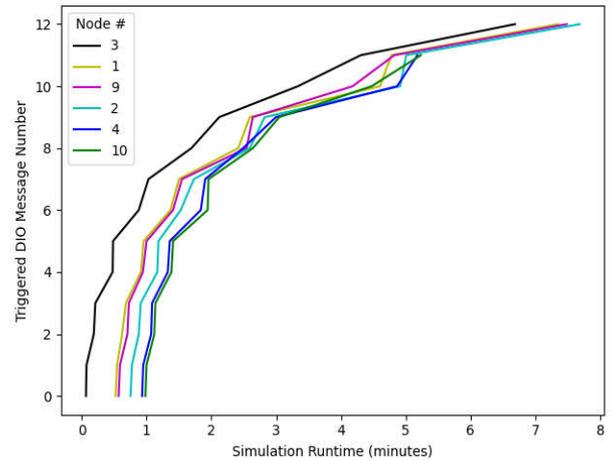

Fig. 6. The time a particular DIO message was triggered for a particular node.

The tenth DIO message was subsequently triggered at approximately the 4-minute mark. That is, it took two minutes to send nine messages and almost an additional two minutes to send just one more message. A similar pattern holds for all of the other nodes. Also, we consider the 3-minute mark as the approximate point in time at which the network reached a steady-state. With the help of this information, we re-ran the simulation a second time and paused it at the 3-minute mark in order to record the initial energy consumption.

*2) Transient State*

On the third minute, we paused the simulation and removed node #10. Then we reset the power tracker so that the measurements would reflect the transient state usage after a node was removed. In order to gather enough data, we let the simulation run for eight minutes. Then we saved the Mote Output into another text file and re-generated the previous graphs.

Figure 7 is the Trickle Timer graph with node #10 removed at the 3-minute mark. As the figure shows, the line for node #10 (purple) stops growing. The lines for nodes #2, #3, and #4 continue to increase, indicating that they were not affected by the removal of node #10. Node #9, on the other hand, has a sharp decrease to its minimum trickle timer value before starting to increase again.

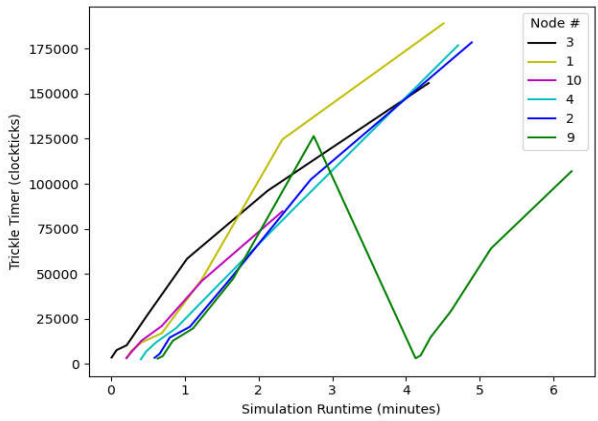

Fig. 7. The Trickle Timer graph with node #10 removed at the 3-minute mark.

Figure 8 shows the triggered DIO messages. The number of DIO messages for node #9 starts to increase sharply from around 4.25 minutes to 6.25 minutes. In that time span, approximately nine DIO messages were triggered for node #9. From 6.25 minutes to eight minutes, only one DIO message was triggered for node #9. Similarly, we consider minute seven as the approximate time at which the network returned to a steady-state from the transient state. With the help of that information, we re-ran the simulation and paused it at minute 3 to remove node #10, and paused it at minute 7 to measure the power consumption of the network during the transient state.

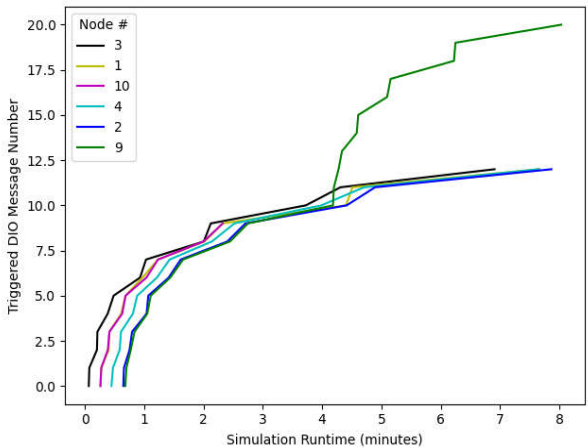

Fig. 8. The triggered DIO messages

### B. Simulation without Orchestra enabled on TSCH

In order to create a basis for comparison, we ran the same simulation without Orchestra enabled for TSCH. The standard scheduling method for TSCH in Cooja is "6TiSCH minimal schedule which emulates an always-on link on top of TSCH. The schedule consists of a single shared slot for all transmissions and receptions in a slot frame" [12].

#### 1) Determining Steady State

Figure 9 and Figure 10 show the result of a simulation that was run without any mote being removed. As Figure 9 shows, the trickle timer continued to increase fairly linearly, indicating that there were no significant changes in the network. As Figure 10 shows, we will consider the 3-minute mark as the approximate point in time at which the network reached a steady-state.

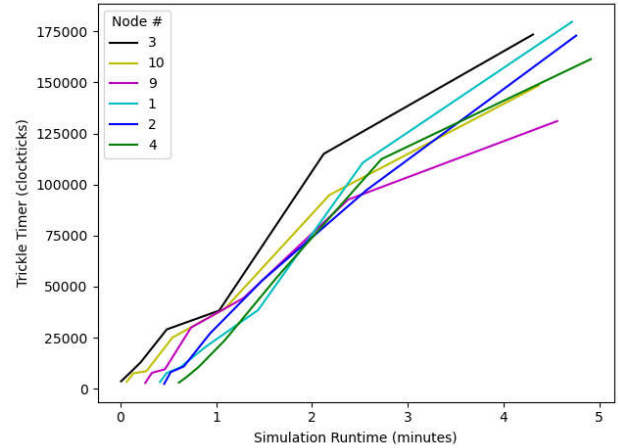

Fig. 9. The trickle timer continued to increase fairly linearly.

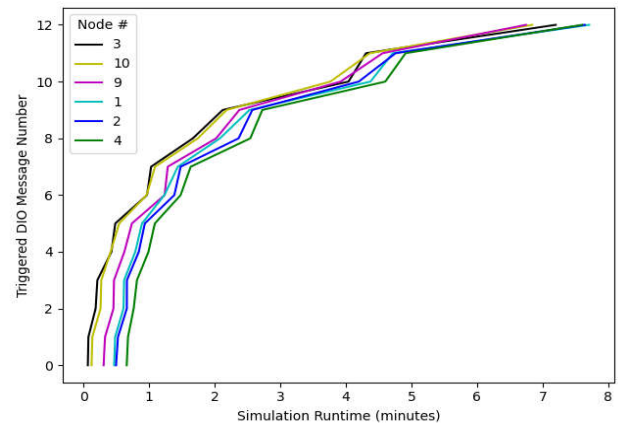

Fig. 10. The approximate time that the network reached a steady-state.

#### 2) Transient State

Figure 11 shows the trickle timer when the simulation was run with mote #10 removed at minute three. The trickle timer line for mote #10 (purple) stopped growing because the mote had died. The neighbors of mote #10, mainly motes #2 and #9 (green and blue, respectively), reset their trickle timers to their minimum values at the 3-minute mark. Interestingly, when running with Orchestra, only node #9 had its trickle timer reset while without Orchestra, both #2 and #9 are reset. After going down to their minimum values, the trickle timers of nodes #2 and #9 began to increase again.

Figure 12 shows the time at which a particular DIO message was triggered for a particular node. As the figure shows, the motes not affected by the removal of mote #10 remain in steady-state. On the other hand, node #2 and #9 begin to increase the number of DIO messages significantly from the 3.5-minute mark until the ~5.5-minute mark. Only one more DIO message is triggered from 5.5-minute mark to 8-minute mark.

Similarly, we consider minute 7 as the approximate time at which the network returned to a steady-state from the transient state. The energy consumed in the transient state, measuring from the third minute when node #10 was removed to minute seven. The average energy consumed (Radio ON) without

Orchestra was 4.23% which is higher than with Orchestra (2.9%).

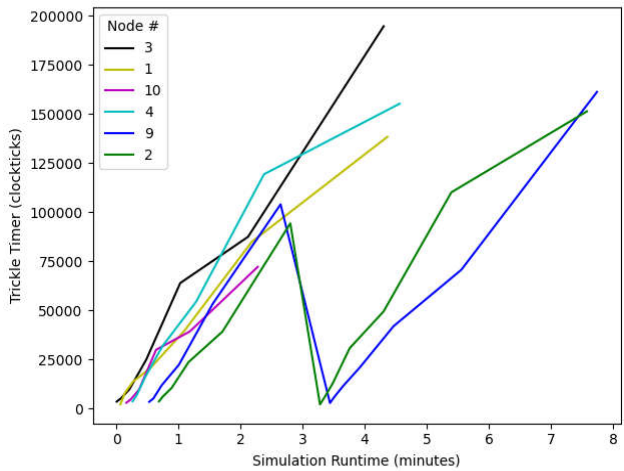

Fig. 11. The trickle timer when the simulation was run with mote #10 removed at minute 3.

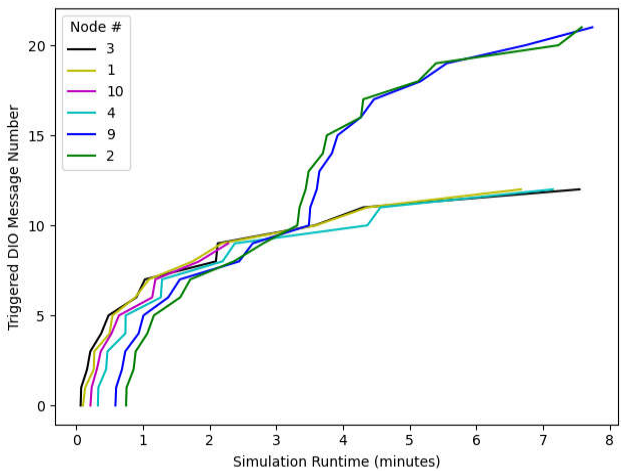

Fig. 12. The time that a particular DIO message was triggered for a particular node.

## VI. CONCLUSION

We used the Cooja simulator to determine how long it takes for a simple network topology to return to a steady-state when a node is removed, as well as how much power is consumed. The experiment was first performed with Orchestra over TSCH and then without Orchestra over TSCH for comparison. We showed that the power consumption with Orchestra would be lower in the transient state because it does not require additional messages in order to create the schedule. The results show the transient state with Orchestra had the "Radio ON" on average for 2.9% of the time, while without Orchestra, the same value was 4.3%. The average time it took for the network to re-stabilize from the 3-minute mark, when a node was removed, was approximately four minutes with and without Orchestra.